%% file: qapl2013.tex
\documentclass[copyright]{eptcs}

\providecommand{\event}{QAPL 2013}

\providecommand{\firstpage}{1}

\usepackage{amsmath,
	    amsfonts,
	    amstext,
	    amssymb,
	    mathrsfs,
	    stmaryrd}
\usepackage{latexsym}
\usepackage{graphicx}
\usepackage{epsf}
\usepackage{greek}
\input{Commands}

\input{Environments}

\title{The Spectrum of Strong Behavioral Equivalences \\
       for Nondeterministic and Probabilistic Processes}

\author{Marco Bernardo
\institute{Dipartimento di Scienze di Base e Fondamenti -- Universit\`a di Urbino -- Italy}
\and
Rocco De Nicola
\institute{IMT -- Institute for Advanced Studies Lucca -- Italy}
\and
Michele Loreti
\institute{Dipartimento di Statistica, Informatica, Applicazioni -- Universit\`a di Firenze -- Italy}}

\def\titlerunning{The Spectrum of Behavioral Equivalences for Nondeterministic and Probabilistic Processes}
\def\authorrunning{M.~Bernardo, R.~De~Nicola, and M.~Loreti}

\begin{document}

\maketitle


\begin{abstract}
We present a spectrum of trace-based, testing, and bisimulation equivalences for nondeterministic and
probabilistic processes whose activities are all observable. For every equivalence under study, we examine
the discriminating power of three variants stemming from three approaches that differ for the way
probabilities of events are compared when nondeterministic choices are resolved via deterministic
schedulers. We show that the first approach -- which compares two resolutions relatively to the probability
distributions of all considered events -- results in a fragment of the spectrum compatible with the spectrum
of behavioral equivalences for fully probabilistic processes. In contrast, the second approach -- which
compares the probabilities of the events of a resolution with the probabilities of the same events in
possibly different resolutions -- gives rise to another fragment composed of coarser equivalences that
exhibits several analogies with the spectrum of behavioral equivalences for fully nondeterministic
processes. Finally, the third approach -- which only compares the extremal probabilities of each event
stemming from the different resolutions -- yields even coarser equivalences that, however, give rise to a
hierarchy similar to that stemming from the second approach.
\end{abstract}


%
%
\section{Introduction}\label{sec:intro}
%
%

Process algebras are mathematically rigorous languages that have been widely used to model and analyze the
behavior of interacting systems. Their structural operational semantics associates with each process term a
labeled transition system (LTS), whose states are the terms themselves and whose labels are the actions that
each term can perform. In order to abstract from unwanted details, the operational semantics is often
coupled with observational mechanisms that permit equating those systems that cannot be distinguished by
external entities. The resulting behavioral equivalences heavily depend on how the specified systems are
expected to be used. Indeed, there is still disagreement on which are the ``reasonable'' observations and
how their outcomes can be used to distinguish or identify systems. Thus, many equivalences have been
proposed and much work has been done to assess their discriminating power and mutual relationships.

The first study in this direction was done by~\cite{DeN87}. There, most of the then known equivalences over
LTS models were ``ordered'' and it was shown that \emph{trace equivalences} (equating systems performing the
same sequences of actions) are strictly coarser than \emph{decorated-trace equivalences} (equating systems
performing the same sequences of actions and refusing/accepting the same sets of actions after them), which
in turn are strictly coarser than \emph{bisimulation equivalences} (equating systems performing the same
sequences of actions and recursively exhibiting the same behavior). It was also shown that the equivalence
obtained by \emph{testing} processes with external observers was coincident with \emph{failure equivalence}
obtained via traces decorated with refusal sets. Afterwards, \cite{Gla01} built the first spectrum that
relates twelve different equivalences and set up a general testing scenario that could be used to generate
many more equivalences.

When process algebras have been enriched with additional dimensions to deal with probabilistic, stochastic,
and timed systems, new behavioral equivalences have been defined and possible classifications have been
proposed. Here, we would like to concentrate on equivalences for probabilistic systems. For this class of
systems, comparative results have been obtained only for so-called fully probabilistic
systems~\cite{JS90,HT92,BKHW05} or only for bisimulation and testing
relations~\cite{BKHW05,LN04,ST05,Wol05}.

In this paper, we aim at a systematic account of the known probabilistic equivalences for nondeterministic
\emph{and} probabilistic systems and introduce, motivate, and relate some new ones. We shall consider an
extension of the LTS model combining nondeterminism and probability that we call NPLTS, in which every
action-labeled transition goes from a source state to a probability distribution over target states rather
than to a single target state~\cite{LS91,Seg95a}. Actions will be assumed to be visible (i.e., we shall not
admit $\tau$-actions) and, for the considered strong equivalences, resolutions of nondeterminism will be
derived by applying memoryless deterministic (as opposed to randomized) schedulers.

When defining behavioral relations over NPLTS models, the idea is to compare resolutions on the basis of the
probabilities of \emph{equivalence-specific events}, like (i)~performing certain sequences of actions,
(ii)~exhibiting certain decorated traces, or (iii)~reaching certain sets of equivalent states via given
actions.

The typical approach followed in the literature (see, e.g., \cite{SL94,Seg95b,Seg96}) consists of comparing
the \emph{probability distributions of all equivalence-specific events} of two resolutions. Two processes
are considered as equivalent if, for each resolution of any of the two processes, there exists a resolution
of the other process such that the probability of \emph{each} equivalence-specific event is the same in the
two resolutions (\emph{fully matching resolutions}). For the known relations based on this approach, we have
that the probabilistic bisimilarity in~\cite{SL94} implies the probabilistic failure equivalence
in~\cite{Seg96} that in turn implies the probabilistic trace equivalence in~\cite{Seg95b}. All these
relations are conservative extensions of the corresponding relations defined over fully nondeterministic
models~\cite{HM85,BHR84} and fully probabilistic models~\cite{GJS90,JS90,HT92}, but in many situations they
turn out to have a high discriminating power.

A different approach has been followed in the literature for defining testing equivalences (see, e.g.,
\cite{YL92,JY95,Seg96,DGHM08}). Instead of comparing individual resolutions of the parallel composition of
processes and tests, the comparison is performed between the \emph{extremal probabilities} of reaching
success \emph{over all resolutions} generated by the experiments on processes under test
(\emph{max-min-matching resolution sets}). In this case, it holds that the resulting probabilistic testing
equivalence is implied by the probabilistic bisimilarity in~\cite{SL94}, but it is related neither to the
probabilistic failure equivalence in~\cite{Seg96} nor to the probabilistic trace equivalence
in~\cite{Seg95b} when restricting attention to deterministic schedulers. Moreover, the resulting
probabilistic testing equivalence subsumes testing equivalence for fully probabilistic
processes~\cite{CDSY99}, but it is not a conservative extension of testing equivalence for fully
nondeterministic processes~\cite{DH84}.

Recently, in~\cite{DMRS08,TDZ11,SZG11,BDL12,BDL13c} a further approach has appeared that compares resolutions
on the basis of the \emph{probabilities of individual equivalence-specific events}. Thus, a resolution of
any of the two processes can be matched, with respect to \emph{different equivalence-specific events}, by
\emph{different resolutions} of the other process (\emph{partially matching resolutions}). For the
behavioral relations resulting from this approach, which weakens the impact of schedulers, we have that
probabilistic bisimilarity implies probabilistic failure equivalence, which in turn implies probabilistic
testing equivalence, which finally implies probabilistic trace equivalence. This approach has contributed to
the development of new probabilistic bisimilarities in~\cite{DMRS08}, \cite{TDZ11,BDL13c}, and~\cite{SZG11}
that, unlike the one in~\cite{SL94}, are characterized by standard probabilistic logics such as quantitative
$\mu$-calculus, PML, and PCTL/PCTL*, respectively. Moreover, in the case of testing equivalence this
approach has the advantage of being conservative also for fully nondeterministic models~\cite{BDL12}, while
in the case of trace equivalence it surprisingly results in a congruence with respect to parallel
composition (full version of \cite{BDL12}).

In our view, the motivations behind the three approaches outlined above are all very reasonable. Indeed,
when applied to fully nondeterministic processes or fully probabilistic processes, they give rise to
well-studied relations that for the fully nondeterministic setting fit into the spectra
in~\cite{DeN87,Gla01} and for the fully probabilistic setting fit into the spectra in~\cite{JS90,HT92}. The
situation is significantly different when the three approaches are instantiated for nondeterministic
\emph{and} probabilistic processes, as in that case they give rise to a much wider variety of relations.

In this paper, we study the relationships between the equivalences for NPLTS models that stem from the three
approaches. For each approach, we consider the three main families of equivalences, namely trace-based,
testing, and bisimulation equivalences. To the best of our knowledge, this is the first comparative study of
different kinds of behavioral equivalences over models featuring both nondeterministic and probabilistic
aspects. Such a study is even more on demand after the recent introduction of new equivalences, like the
ones in~\cite{DMRS08,TDZ11,SZG11,BDL12,BDL13c}, that have interesting properties.

To have a full picture of the spectrum, the reader is referred to Fig.~\ref{fig:spectrum} in the concluding
section. There, the equivalences stemming from the same approach are contained in boxes with the same shape
(hexagonal, rounded, or rectangular) and the equivalences specifically introduced for the purposes of this
paper are in dashed boxes. We would like to stress that the original contribution of the paper is not given
by the equivalences that we introduce to fill in gaps, \emph{but is the spectrum itself}. 

We shall see that the family of equivalences that assign a central role to schedulers by requiring fully
matching resolutions, yields a hierarchy that is in accordance with the one for fully probabilistic
processes in~\cite{JS90,HT92}. Conversely, the family of equivalences that assign a weaker role to
schedulers by requiring partially matching resolutions, gives rise to relations that are coarser than the
former and yields a hierarchy that is in accordance with the one for fully nondeterministic processes
in~\cite{DeN87,Gla01}. Finally, the family of equivalences that only consider extremal probabilities, has
again several analogies with the fully nondeterministic spectrum and yields even coarser equivalences. There
are however some noticeable anomalies in the last two families, given by a few equivalences suffering from
isolation.

The rest of the paper is organized as follows. In Sect.~\ref{sec:nplts}, we introduce the NPLTS model. In
Sects.~\ref{sec:trace_equiv} to~\ref{sec:bisim_equiv}, we define and compare, respectively, the trace-based,
testing, and bisimulation equivalences that arise from the three approaches outlined above. Finally, in
Sect.~\ref{sec:concl} we draw some conclusions and graphically summarize the results by depicting the
spectrum of all the considered equivalences.

%
%
\section{Nondeterministic and Probabilistic Processes}\label{sec:nplts}
%
%

Processes combining nondeterminism and probability are typically described by means of extensions of the LTS
model, in which every action-labeled transition goes from a source state to a \emph{probability distribution
over target states} rather than to a single target state. They are essentially Markov decision processes and
are representative of a number of slightly different probabilistic computational models including internal
nondeterminism such as, e.g., concurrent Markov chains~\cite{Var85}, alternating probabilistic
models~\cite{HJ90,YL92,PLS00}, probabilistic automata in the sense of~\cite{Seg95a}, and the denotational
probabilistic models in~\cite{JSM97} (see~\cite{SD04} for an overview). We formalize them as a variant of
simple probabilistic automata~\cite{Seg95a}.

	\begin{definition}\label{def:nplts}

A nondeterministic and probabilistic labeled transition system, NPLTS for short, is a triple $(S, A, \!
\arrow{}{} \!)$ where $S$ is an at most countable set of states, $A$ is a countable set of
transition-labeling actions, and $\! \arrow{}{} \! \subseteq S \times A \times \ms{Distr}(S)$ is a
transition relation with $\ms{Distr}(S)$ being the set of discrete probability distributions over $S$.
\fullbox

	\end{definition}

A transition $(s, a, \cald)$ is written $s \arrow{a}{} \cald$. We say that $s' \in S$ is not reachable from
$s$ via that \linebreak $a$-transition if $\cald(s') = 0$, otherwise we say that it is reachable with
probability $p = \cald(s')$. The reachable states form the support of $\cald$: $\ms{supp}(\cald) = \{ s' \in
S \mid \cald(s') > 0 \}$. We write $s \arrow{a}{} \!$ to indicate that $s$ has an $a$-transition. The choice
among all the transitions departing from $s$ is nondeterministic, while the choice of the target state for a
specific transition is probabilistic. An NPLTS represents (i)~a \emph{fully nondeterministic process} when
every transition leads to a distribution that concentrates all the probability mass into a single state or
(ii) a \emph{fully probabilistic process} when every state has at most one outgoing transition.

An NPLTS can be depicted as a directed graph-like structure in which vertices represent states and
action-labeled edges represent action-labeled transitions. Given a transition $s \arrow{a}{} \cald$, the
corresponding $a$-labeled edge goes from the vertex representing state $s$ to a set of vertices linked by a
dashed line, each of which represents a state $s' \in \ms{supp}(\cald)$ and is labeled with $\cald(s')$ --
label omitted if $\cald(s') = 1$. Figure~\ref{fig:counterex_trace} shows eighteen NPLTS models, nine of
which are fully nondeterministic.

In this setting, a computation is a sequence of state-to-state steps, each denoted by $s \step{a}{} s'$ and
derived from a state-to-distribution transition. Formally, given an NPLTS $\call = (S, A, \! \arrow{}{} \!)$
and $s, s' \in S$, we say that $c \: \equiv \: s_{0} \step{a_{1}}{} s_{1} \step{a_{2}}{} s_{2} \dots s_{n -
1} \step{a_{n}}{} s_{n}$ is a computation of $\call$ of length $n$ from $s = s_{0}$ to $s' = s_{n}$ iff for
all $i = 1, \dots, n$ there exists a transition $s_{i - 1} \arrow{a_{i}}{} \cald_{i}$ such that $s_{i} \in
\ms{supp}(\cald_{i})$, with $\cald_{i}(s_{i})$ being the execution probability of step $s_{i - 1}
\step{a_{i}}{} s_{i}$ conditioned on the selection of transition $s_{i - 1} \arrow{a_{i}}{} \cald_{i}$ of
$\call$ at state~$s_{i - 1}$. \linebreak We denote by $\ms{first}(c)$ and $\ms{last}(c)$ the initial state
and the final state of $c$, respectively, and by $\calc_{\rm fin}(s)$ the set of finite-length computations
from $s$.

We call resolution of $s$ any possible way of resolving nondeterminism starting from $s$. Each resolution is
a tree-like structure whose branching points represent probabilistic choices. This is obtained by unfolding
from $s$ the graph structure underlying~$\call$ and by selecting at each state a single transition
of~$\call$ (\emph{deterministic scheduler}) or a convex combination of equally labeled transitions of
$\call$ (\emph{randomized scheduler}) among all the transitions possible from that state. Below, we
introduce the notion of resolution arising from a deterministic scheduler as a fully probabilistic NPLTS.
Notice that, when $\call$ is fully nondeterministic, resolutions boil down to computations.

	\begin{definition}\label{def:resolution}

Let $\call = (S, A, \! \arrow{}{} \!)$ be an NPLTS and $s \in S$. We say that an NPLTS $\calz = (Z, A, \!
\arrow{}{\calz} \!)$ is a resolution of~$s$ obtained via a deterministic scheduler iff there exists a state
correspondence function $\ms{corr}_{\calz} : Z \rightarrow S$ such that $s = \ms{corr}_{\calz}(z_{s})$, for
some $z_{s} \in Z$, and for all $z \in Z$ it holds that:

		\begin{itemize}

\item If $z \arrow{a}{\calz} \cald$, then $\ms{corr}_{\calz}(z) \arrow{a}{} \cald'$ with $\cald(z') =
\cald'(\ms{corr}_{\calz}(z'))$ for all $z' \in Z$.

\item If $z \arrow{a_{1}}{\calz} \cald_{1}$ and $z \arrow{a_{2}}{\calz} \cald_{2}$, then $a_{1} = a_{2}$ and
$\cald_{1} = \cald_{2}$.
\fullbox

		\end{itemize}

	\end{definition}

We denote by $\ms{Res}(s)$ the set of resolutions of~$s$ and by $\ms{Res}_{\rm max}(s)$ the set of maximal
resolutions of~$s$, i.e., the resolutions of~$s$ that cannot be further extended in accordance with the
graph structure of $\call$ and the constraints above. Since $\calz \in \ms{Res}(s)$ is fully probabilistic,
the probability $\ms{prob}(c)$ of executing $c \in \calc_{\rm fin}(z_{s})$ can be defined as the product of
the (no longer conditional) execution probabilities of the individual steps of $c$, with $\ms{prob}(c)$
being always equal to $1$ if $\call$ is fully nondeterministic. This notion is lifted to $C \subseteq
\calc_{\rm fin}(z_{s})$ by letting $\ms{prob}(C) = \sum_{c \in C} \ms{prob}(c)$ whenever none of the
computations in $C$ is a proper prefix of one of the others.

We finally introduce a notion of fully synchronous parallel composition for NPLTS models that is
instrumental to the definition of testing equivalences.

	\begin{definition}\label{def:parcomp}

Let $\call_{i} = (S_{i}, A, \! \arrow{}{i} \!)$ be an NPLTS for $i = 1, 2$. The parallel composition of
$\call_{1}$ and $\call_{2}$ is the NPLTS $\call_{1} \pco{} \call_{2} = (S_{1} \times S_{2}, A, \! \arrow{}{}
\!)$ where $\! \arrow{}{} \! \subseteq (S_{1} \times S_{2}) \times A \times \ms{Distr}(S_{1} \times S_{2})$
is such that $(s_{1}, s_{2}) \arrow{a}{} \cald$ iff $s_{1} \arrow{a}{1} \cald_{1}$ and $s_{2} \arrow{a}{2}
\cald_{2}$ with $\cald(s'_{1}, s'_{2}) = \cald_{1}(s'_{1}) \cdot \cald_{2}(s'_{2})$ for each $(s'_{1},
s'_{2}) \in S_{1} \times S_{2}$.
\fullbox

	\end{definition}

%
%
\section{Trace-Based Equivalences for NPLTS Models}\label{sec:trace_equiv}
%
%

Trace-based equivalences examine the probability with which two states perform computations labeled with the
same (decorated) traces for each possible way of resolving nondeterminism. As outlined in
Sect.~\ref{sec:intro}, there are three different approaches to defining them. The first approach is to match
resolutions according to \emph{trace-based distributions}, which means that for each resolution of one of
the two states there must exist a resolution of the other state such that, \emph{for every (decorated)
trace}, the two resolutions have the same probability of performing a computation labeled with that
(decorated) trace. In other words, matching resolutions of the two states are related by the fully
probabilistic version of the trace-based equivalence (fully matching resolutions). The second approach is to
consider \emph{a single (decorated) trace at a time}, i.e., to anticipate the quantification over
(decorated) traces with respect to the quantification over resolutions. In this way, differently labeled
computations of a resolution of one of the two states are allowed to be matched by computations of several
different resolutions of the other state (partially matching resolutions). The third approach is to compare
only the \emph{extremal probabilities} of performing each (decorated) trace over the various resolutions
(max-min-matching resolution sets).

We say that a computation is compatible with a trace $\alpha \in A^{*}$ iff the sequence of actions labeling
its steps is equal to~$\alpha$. Given an NPLTS $\call = (S, A, \! \arrow{}{} \!)$, $s \in S$, and $\calz \in
\ms{Res}(s)$, we denote by $\calcc(z_{s}, \alpha)$ the set of $\alpha$-compatible computations in
$\calc_{\rm fin}(z_{s})$ and by $\ms{Res}_{\alpha}(s)$ the set of resolutions in $\ms{Res}(s)$ having no
computations corresponding to proper prefixes of $\alpha$-compatible computations of~$\call$. In each of the
following definitions, we assume $s_{1}, s_{2} \in S$ and we explicitly add a reference whenever the defined
equivalence has already appeared in the literature. In some definitions, we indicate with $\sqcup$/$\sqcap$
the supremum/infimum of a set of numbers in $\realns_{[0, 1]}$ and we assume it to be~$0$ when the set is
empty.

	\begin{definition}\label{def:ptrdis} 

(\emph{Probabilistic trace-distribution equivalence} -- $\sbis{\rm PTr,dis}$ -- \cite{Seg95b}) \\
$s_{1} \sbis{\rm PTr,dis} s_{2}$ iff for each $\calz_{1} \in \ms{Res}(s_{1})$ there exists $\calz_{2} \in
\ms{Res}(s_{2})$ such that \underline{for all $\alpha \in A^{*}$}:
\cws{0}{\ms{prob}(\calcc(z_{s_{1}}, \alpha)) \: = \: \ms{prob}(\calcc(z_{s_{2}}, \alpha))}
and symmetrically for each $\calz_{2} \in \ms{Res}(s_{2})$.
\fullbox

	\end{definition}

	\begin{definition}\label{def:ptr}

(\emph{Probabilistic trace equivalence} -- $\sbis{\rm PTr}$ -- \cite{BDL12}) \\
$s_{1} \sbis{\rm PTr} s_{2}$ iff \underline{for all $\alpha \in A^{*}$} it holds that for each $\calz_{1}
\in \ms{Res}(s_{1})$ there exists $\calz_{2} \in \ms{Res}(s_{2})$ such that:
\cws{0}{\ms{prob}(\calcc(z_{s_{1}}, \alpha)) \: = \: \ms{prob}(\calcc(z_{s_{2}}, \alpha))}
and symmetrically for each $\calz_{2} \in \ms{Res}(s_{2})$.
\fullbox

	\end{definition}

	\begin{definition}\label{def:ptrsupinf}

(\emph{Probabilistic $\sqcup\sqcap$-trace equivalence} -- $\sbis{\rm PTr,\sqcup\sqcap}$) \\
$s_{1} \sbis{\rm PTr,\sqcup\sqcap} s_{2}$ iff for all $\alpha \in A^{*}$:
\cws{10}{\begin{array}{rcl}
\bigsqcup\limits_{\calz_{1} \in \ms{Res}_{\alpha}(s_{1})} \ms{prob}(\calcc(z_{s_{1}}, \alpha)) & \!\!\! =
\!\!\! & \bigsqcup\limits_{\calz_{2} \in \ms{Res}_{\alpha}(s_{2})} \ms{prob}(\calcc(z_{s_{2}}, \alpha))
\\[0.4cm]
\bigsqcap\limits_{\calz_{1} \in \ms{Res}_{\alpha}(s_{1})} \ms{prob}(\calcc(z_{s_{1}}, \alpha)) & \!\!\! =
\!\!\! & \bigsqcap\limits_{\calz_{2} \in \ms{Res}_{\alpha}(s_{2})} \ms{prob}(\calcc(z_{s_{2}}, \alpha)) \\
\end{array}}
\fullbox

	\end{definition}

A variant that additionally considers completed computations was introduced in the literature of fully
nondeterministic models in order to equip trace equivalence with deadlock sensitivity. We denote by
$\calccc(z_{s}, \alpha)$ the set of completed $\alpha$-compatible computations from $z_{s}$. Each of these
computations $c$ belongs to $\calcc(z_{s}, \alpha)$ and is such that $\ms{corr}_{\calz}(\ms{last}(c))$ has
no outgoing transitions in~$\call$.

	\begin{definition}\label{def:pctrdis} 

(\emph{Probabilistic completed-trace-distribution equivalence} -- $\sbis{\rm PCTr,dis}$) \\
$s_{1} \sbis{\rm PCTr,dis} s_{2}$ iff for each $\calz_{1} \in \ms{Res}(s_{1})$ there exist $\calz_{2},
\calz'_{2} \in \ms{Res}(s_{2})$ such that \underline{for all $\alpha \in A^{*}$}:
\cws{0}{\begin{array}{rcl}
\ms{prob}(\calcc(z_{s_{1}}, \alpha)) & \!\!\! = \!\!\! & \ms{prob}(\calcc(z_{s_{2}}, \alpha)) \\
\ms{prob}(\calccc(z_{s_{1}}, \alpha)) & \!\!\! = \!\!\! & \ms{prob}(\calccc(z'_{s_{2}}, \alpha)) \\
\end{array}}
and symmetrically for each $\calz_{2} \in \ms{Res}(s_{2})$.
\fullbox

	\end{definition}

	\begin{definition}\label{def:pctr}

(\emph{Probabilistic completed-trace equivalence} -- $\sbis{\rm PCTr}$) \\
$s_{1} \sbis{\rm PCTr} s_{2}$ iff \underline{for all $\alpha \in A^{*}$} it holds that for each $\calz_{1}
\in \ms{Res}(s_{1})$ there exist $\calz_{2}, \calz'_{2} \in \ms{Res}(s_{2})$ such that:
\cws{0}{\begin{array}{rcl}
\ms{prob}(\calcc(z_{s_{1}}, \alpha)) & \!\!\! = \!\!\! & \ms{prob}(\calcc(z_{s_{2}}, \alpha)) \\
\ms{prob}(\calccc(z_{s_{1}}, \alpha)) & \!\!\! = \!\!\! & \ms{prob}(\calccc(z'_{s_{2}}, \alpha)) \\
\end{array}}
and symmetrically for each $\calz_{2} \in \ms{Res}(s_{2})$.
\fullbox

	\end{definition}

	\begin{definition}\label{def:pctrsupinf}

(\emph{Probabilistic $\sqcup\sqcap$-completed-trace equivalence} -- $\sbis{\rm PCTr,\sqcup\sqcap}$) \\
$s_{1} \sbis{\rm PCTr,\sqcup\sqcap} s_{2}$ iff for all $\alpha \in A^{*}$:
\cws{0}{\begin{array}{rcl}
\bigsqcup\limits_{\calz_{1} \in \ms{Res}_{\alpha}(s_{1})} \ms{prob}(\calcc(z_{s_{1}}, \alpha)) & \!\!\! =
\!\!\! & \bigsqcup\limits_{\calz_{2} \in \ms{Res}_{\alpha}(s_{2})} \ms{prob}(\calcc(z_{s_{2}}, \alpha))
\\[0.4cm]
\bigsqcap\limits_{\calz_{1} \in \ms{Res}_{\alpha}(s_{1})} \ms{prob}(\calcc(z_{s_{1}}, \alpha)) & \!\!\! =
\!\!\! & \bigsqcap\limits_{\calz_{2} \in \ms{Res}_{\alpha}(s_{2})} \ms{prob}(\calcc(z_{s_{2}}, \alpha)) \\
\end{array}}
and:
\cws{10}{\begin{array}{rcl}
\bigsqcup\limits_{\calz_{1} \in \ms{Res}_{\alpha}(s_{1})} \ms{prob}(\calccc(z_{s_{1}}, \alpha)) & \!\!\! =
\!\!\! & \bigsqcup\limits_{\calz_{2} \in \ms{Res}_{\alpha}(s_{2})} \ms{prob}(\calccc(z_{s_{2}}, \alpha))
\\[0.4cm]
\bigsqcap\limits_{\calz_{1} \in \ms{Res}_{\alpha}(s_{1})} \ms{prob}(\calccc(z_{s_{1}}, \alpha)) & \!\!\! =
\!\!\! & \bigsqcap\limits_{\calz_{2} \in \ms{Res}_{\alpha}(s_{2})} \ms{prob}(\calccc(z_{s_{2}}, \alpha)) \\
\end{array}}
\fullbox

	\end{definition}

Failure semantics generalizes completed-trace equivalence towards arbitrary safety properties. A failure
pair is an element $\varphi \in A^{*} \times 2^{A}$ formed by a trace $\alpha$ and a decoration~$F$ called
failure set. We say that $c \in \calc_{\rm fin}(z_{s})$ is compatible with $\varphi$ iff $c \in
\calcc(z_{s}, \alpha)$ and $\ms{corr}_{\calz}(\ms{last}(c))$ has no outgoing transitions in~$\call$ labeled
with an action in $F$. We denote by $\calfcc(z_{s}, \varphi)$ the set of $\varphi$-compatible computations
from~$z_{s}$. Moreover, we call failure trace an element $\phi \in (A \times 2^{A})^{*}$ given by a sequence
of $n \in \natns$ pairs of the form $(a_{i}, F_{i})$. We say that $c \in \calc_{\rm fin}(z_{s})$ is
compatible with $\phi$ iff $c \in \calcc(z_{s}, a_{1} \dots a_{n})$ and, denoting by $z_{i}$ the state
reached by~$c$ after the $i$-th step for all $i = 1, \dots, n$, $\ms{corr}_{\calz}(z_{i})$ has no outgoing
transitions in $\call$ labeled with an action in $F_{i}$. We denote by $\calftcc(z_{s}, \phi)$ the set of
$\phi$-compatible computations from $z_{s}$.

	\begin{definition}\label{def:pfdis}

(\emph{Probabilistic failure-distribution equivalence} -- $\sbis{\rm PF,dis}$ -- \cite{Seg96}) \\
$s_{1} \sbis{\rm PF,dis} s_{2}$ iff for each $\calz_{1} \in \ms{Res}(s_{1})$ there exists $\calz_{2} \in
\ms{Res}(s_{2})$ such that \underline{for all $\varphi \in A^{*} \times 2^{A}$}:
\cws{0}{\ms{prob}(\calfcc(z_{s_{1}}, \varphi)) \: = \: \ms{prob}(\calfcc(z_{s_{2}}, \varphi))}
and symmetrically for each $\calz_{2} \in \ms{Res}(s_{2})$.
\fullbox

	\end{definition}

	\begin{definition}\label{def:pf}

(\emph{Probabilistic failure equivalence} -- $\sbis{\rm PF}$ -- \cite{BDL12}) \\
$s_{1} \sbis{\rm PF} s_{2}$ iff \underline{for all $\varphi \in A^{*} \times 2^{A}$} it holds that for each
$\calz_{1} \in \ms{Res}(s_{1})$ there exists $\calz_{2} \in \ms{Res}(s_{2})$ such that:
\cws{0}{\ms{prob}(\calfcc(z_{s_{1}}, \varphi)) \: = \: \ms{prob}(\calfcc(z_{s_{2}}, \varphi))}
and symmetrically for each $\calz_{2} \in \ms{Res}(s_{2})$.
\fullbox

	\end{definition}

	\begin{definition}\label{def:pfsupinf}

(\emph{Probabilistic $\sqcup\sqcap$-failure equivalence} -- $\sbis{\rm PF,\sqcup\sqcap}$) \\
$s_{1} \sbis{\rm PF,\sqcup\sqcap} s_{2}$ iff for all $\varphi = (\alpha, F) \in A^{*} \times 2^{A}$:
\cws{10}{\begin{array}{rcl}
\bigsqcup\limits_{\calz_{1} \in \ms{Res}_{\alpha}(s_{1})} \ms{prob}(\calfcc(z_{s_{1}}, \varphi)) & \!\!\! =
\!\!\! & \bigsqcup\limits_{\calz_{2} \in \ms{Res}_{\alpha}(s_{2})} \ms{prob}(\calfcc(z_{s_{2}}, \varphi))
\\[0.4cm]
\bigsqcap\limits_{\calz_{1} \in \ms{Res}_{\alpha}(s_{1})} \ms{prob}(\calfcc(z_{s_{1}}, \varphi)) & \!\!\! =
\!\!\! & \bigsqcap\limits_{\calz_{2} \in \ms{Res}_{\alpha}(s_{2})} \ms{prob}(\calfcc(z_{s_{2}}, \varphi)) \\
\end{array}}
\fullbox

	\end{definition}

	\begin{definition}\label{def:pftrdis}

(\emph{Probabilistic failure-trace-distribution equivalence} -- $\sbis{\rm PFTr,dis}$) \\
Same as Def.~\ref{def:pfdis} with $\phi \in (A \times 2^{A})^{*}$ and $\calftcc$ in place of $\varphi \in
A^{*} \times 2^{A}$ and $\calfcc$, respectively.
\fullbox

	\end{definition}

	\begin{definition}\label{def:pftr}

(\emph{Probabilistic failure-trace equivalence} -- $\sbis{\rm PFTr}$) \\
Same as Def.~\ref{def:pf} with $\phi \in (A \times 2^{A})^{*}$ and $\calftcc$ in place of $\varphi \in A^{*}
\times 2^{A}$ and $\calfcc$, respectively.
\fullbox

	\end{definition}

	\begin{definition}\label{def:pftrsupinf}

(\emph{Probabilistic $\sqcup\sqcap$-failure-trace equivalence} -- $\sbis{\rm PFTr,\sqcup\sqcap}$) \\
Same as Def.~\ref{def:pfsupinf} with $\phi \in (A \times 2^{A})^{*}$ and $\calftcc$ in place of $\varphi \in
A^{*} \times 2^{A}$ and $\calfcc$, respectively.
\fullbox

	\end{definition}

A different generalization towards liveness properties is readiness semantics. A ready pair is an element
$\varrho \in A^{*} \times 2^{A}$ formed by a trace $\alpha$ and a decoration~$R$ called ready set. We say
that $c$ is compatible with~$\varrho$ iff $c \in \calcc(z_{s}, \alpha)$ and the set of actions labeling the
transitions in $\call$ departing from $\ms{corr}_{\calz}(\ms{last}(c))$ is precisely~$R$. We denote by
$\calrcc(z_{s}, \varrho)$ the set of $\varrho$-compatible computations from $z_{s}$. Moreover, we call ready
trace an element $\rho \in (A \times 2^{A})^{*}$ given by a sequence of $n \in \natns$ pairs of the form
$(a_{i}, R_{i})$. We say that $c \in \calc_{\rm fin}(z_{s})$ is compatible with $\rho$ iff $c \in
\calcc(z_{s}, a_{1} \dots a_{n})$ and, denoting by $z_{i}$ the state reached by~$c$ after the $i$-th step
for all $i = 1, \dots, n$, the set of actions labeling the transitions in $\call$ departing from
$\ms{corr}_{\calz}(z_{i})$ is precisely $R_{i}$. We denote by $\calrtcc(z_{s}, \rho)$ the set of
$\rho$-compatible computations from $z_{s}$.

	\begin{definition}\label{def:prdis}

(\emph{Probabilistic readiness-distribution equivalence} -- $\sbis{\rm PR,dis}$) \\
$s_{1} \sbis{\rm PR,dis} s_{2}$ iff for each $\calz_{1} \in \ms{Res}(s_{1})$ there exists $\calz_{2} \in
\ms{Res}(s_{2})$ such that \underline{for all $\varrho \in A^{*} \times 2^{A}$}:
\cws{0}{\ms{prob}(\calrcc(z_{s_{1}}, \varrho)) \: = \: \ms{prob}(\calrcc(z_{s_{2}}, \varrho))}
and symmetrically for each $\calz_{2} \in \ms{Res}(s_{2})$.
\fullbox

	\end{definition}

	\begin{definition}\label{def:pr}

(\emph{Probabilistic readiness equivalence} -- $\sbis{\rm PR}$) \\
$s_{1} \sbis{\rm PR} s_{2}$ iff \underline{for all $\varrho \in A^{*} \times 2^{A}$} it holds that for each
$\calz_{1} \in \ms{Res}(s_{1})$ there exists $\calz_{2} \in \ms{Res}(s_{2})$ such that:
\cws{0}{\ms{prob}(\calrcc(z_{s_{1}}, \varrho)) \: = \: \ms{prob}(\calrcc(z_{s_{2}}, \varrho))}
and symmetrically for each $\calz_{2} \in \ms{Res}(s_{2})$.
\fullbox

	\end{definition}

	\begin{definition}\label{def:prsupinf}

(\emph{Probabilistic $\sqcup\sqcap$-readiness equivalence} -- $\sbis{\rm PR,\sqcup\sqcap}$) \\
$s_{1} \sbis{\rm PR,\sqcup\sqcap} s_{2}$ iff for all $\varrho = (\alpha, R) \in A^{*} \times 2^{A}$:
\cws{10}{\begin{array}{rcl}
\bigsqcup\limits_{\calz_{1} \in \ms{Res}_{\alpha}(s_{1})} \ms{prob}(\calrcc(z_{s_{1}}, \varrho)) & \!\!\! =
\!\!\! & \bigsqcup\limits_{\calz_{2} \in \ms{Res}_{\alpha}(s_{2})} \ms{prob}(\calrcc(z_{s_{2}}, \varrho))
\\[0.4cm]
\bigsqcap\limits_{\calz_{1} \in \ms{Res}_{\alpha}(s_{1})} \ms{prob}(\calrcc(z_{s_{1}}, \varrho)) & \!\!\! =
\!\!\! & \bigsqcap\limits_{\calz_{2} \in \ms{Res}_{\alpha}(s_{2})} \ms{prob}(\calrcc(z_{s_{2}}, \varrho)) \\
\end{array}}
\fullbox

	\end{definition}

	\begin{definition}\label{def:prtrdis}

(\emph{Probabilistic ready-trace-distribution equivalence} -- $\sbis{\rm PRTr,dis}$) \\
Same as Def.~\ref{def:prdis} with $\rho \in (A \times 2^{A})^{*}$ and $\calrtcc$ in place of $\varrho \in
A^{*} \times 2^{A}$ and $\calrcc$, respectively.
\fullbox

	\end{definition}

	\begin{definition}\label{def:prtr}

(\emph{Probabilistic ready-trace equivalence} -- $\sbis{\rm PRTr}$) \\
Same as Def.~\ref{def:pr} with $\rho \in (A \times 2^{A})^{*}$ and $\calrtcc$ in place of $\varrho \in A^{*}
\times 2^{A}$ and $\calrcc$, respectively.
\fullbox

	\end{definition}

	\begin{definition}\label{def:prtrsupinf}

(\emph{Probabilistic $\sqcup\sqcap$-ready-trace equivalence} -- $\sbis{\rm PRTr,\sqcup\sqcap}$) \\
Same as Def.~\ref{def:prsupinf} with $\rho \in (A \times 2^{A})^{*}$ and $\calrtcc$ in place of $\varrho \in
A^{*} \times 2^{A}$ and $\calrcc$, respectively.
\fullbox

	\end{definition}

The eighteen trace-based equivalences defined above are all backward compatible with the corresponding
trace-based equivalences respectively defined in~\cite{BHR84,OH86} for fully nondeterministic processes and
in~\cite{JS90,HT92} for fully probabilistic processes.

	\begin{theorem}\label{thm:trace_compat}

Let $\sigma \in \{ {\rm RTr}, {\rm FTr}, {\rm R}, {\rm F}, {\rm CTr}, {\rm Tr} \}$ with $\sbis{\rm
P\sigma,dis}$, $\sbis{\rm P\sigma}$, and $\sbis{\rm P\sigma,\sqcup\sqcap}$ being the equivalences defined
above, $\sbis{\rm \sigma,fnd}$ being the corresponding equivalence defined for fully nondeterministic
processes, and $\sbis{\rm \sigma,fpr}$ being the corresponding equivalence defined for fully probabilistic
processes. Then:

		\begin{enumerate}

\item $\sbis{\rm P\sigma,dis} \: = \: \sbis{\rm P\sigma} \: = \: \sbis{\rm P\sigma,\sqcup\sqcap} \: = \:
\sbis{\rm \sigma,fnd}$ over fully nondeterministic NPLTS models.

\item $\sbis{\rm P\sigma,dis} \: = \: \sbis{\rm P\sigma} \: = \: \sbis{\rm P\sigma,\sqcup\sqcap} \: = \:
\sbis{\rm \sigma,fpr}$ over fully probabilistic NPLTS models.
\fullbox

		\end{enumerate}

	\end{theorem}

We now investigate the relationships among the eighteen trace-based equivalences. As expected, each
equivalence relying on trace-based distributions is finer than the corresponding equivalence considering a
single (decorated) trace at a time, which in turn is finer than the corresponding equivalence based on
extremal probabilities of (decorated) traces. For the equivalences of the first type, similar to the fully
probabilistic spectrum in~\cite{JS90,HT92} it turns out that the readiness semantics coincides with the
failure semantics. In contrast, for the other two types of equivalences, unlike the fully nondeterministic
spectrum in~\cite{Gla01} no connection can be established between readiness semantics and failure semantics.

	\begin{figure}[p]

\centerline{\includegraphics{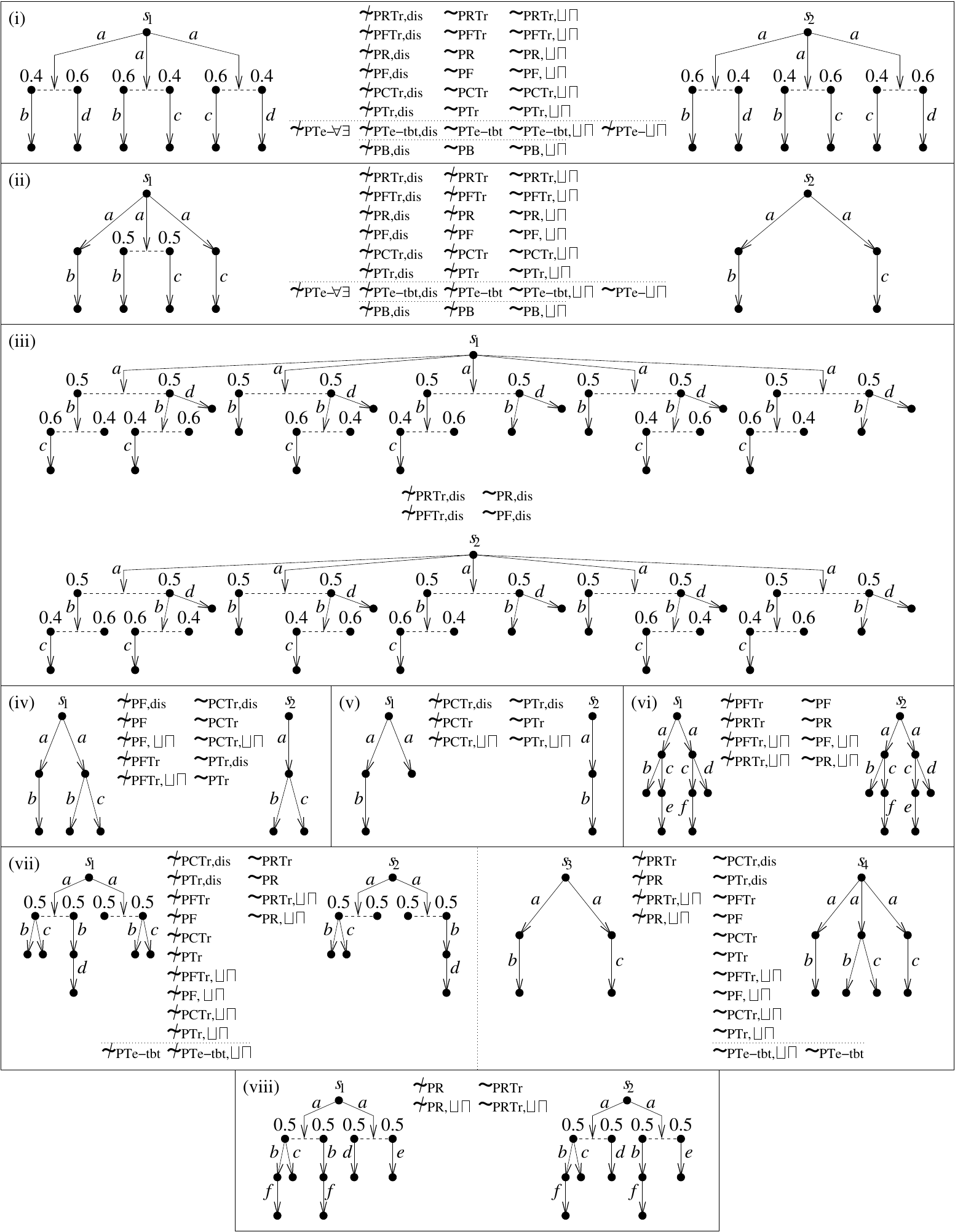}}
\caption{Counterexamples for strict inclusion and incomparability of the trace-based equivalences}
\label{fig:counterex_trace}

	\end{figure}

	\begin{theorem}\label{thm:trace_results}

It holds that:

		\begin{enumerate}

\item $\sbis{\rm \pi,dis} \: \subseteq \: \sbis{\rm \pi} \: \subseteq \: \sbis{\rm \pi,\sqcup\sqcap}$ for
all $\pi \in \{ {\rm PRTr}, {\rm PFTr}, {\rm PR}, {\rm PF}, {\rm PCTr}, {\rm PTr} \}$.

\item $\sbis{\rm PRTr,dis} \: = \: \sbis{\rm PFTr,dis}$ over finitely-branching NPLTS models.

\item $\sbis{\rm PR,dis} \: = \: \sbis{\rm PF,dis}$ over finitely-branching NPLTS models.

\item $\sbis{\rm PFTr,dis} \: \subseteq \: \sbis{\rm PF,dis} \: \subseteq \: \sbis{\rm PCTr,dis} \:
\subseteq \: \sbis{\rm PTr,dis}$.

\item $\sbis{\rm PFTr} \: \subseteq \: \sbis{\rm PF} \: \subseteq \: \sbis{\rm PCTr} \: \subseteq \:
\sbis{\rm PTr}$.

\item $\sbis{\rm PFTr,\sqcup\sqcap} \: \subseteq \: \sbis{\rm PF,\sqcup\sqcap} \: \subseteq \: \sbis{\rm
PCTr,\sqcup\sqcap} \: \subseteq \: \sbis{\rm PTr,\sqcup\sqcap}$.
\fullbox

		\end{enumerate}

	\end{theorem}

All the inclusions above are strict, as shown in Figs.~\ref{fig:counterex_trace}(i) to~(vi). It is worth
noting the isolation of $\sbis{\rm PRTr}$, $\sbis{\rm PR}$, $\sbis{\rm PRTr,\sqcup\sqcap}$, and $\sbis{\rm
PR,\sqcup\sqcap}$, each of which is incomparable with $\sbis{\rm PCTr,dis}$, $\sbis{\rm PTr,dis}$,
$\sbis{\rm PFTr}$, $\sbis{\rm PF}$, $\sbis{\rm PCTr}$, $\sbis{\rm PTr}$, $\sbis{\rm PFTr,\sqcup\sqcap}$,
$\sbis{\rm PF,\sqcup\sqcap}$, $\sbis{\rm PCTr,\sqcup\sqcap}$, and $\sbis{\rm PTr,\sqcup\sqcap}$, as shown in
Fig.~\ref{fig:counterex_trace}(vii). Moreover, Figs.~\ref{fig:counterex_trace}(i) and~(iv) show that
$\sbis{\rm PFTr}$, $\sbis{\rm PF}$, $\sbis{\rm PFTr,\sqcup\sqcap}$, and $\sbis{\rm PF,\sqcup\sqcap}$ are
incomparable with $\sbis{\rm PCTr,dis}$ and $\sbis{\rm PTr,dis}$, while Figs.~\ref{fig:counterex_trace}(ii)
and~(iv) show that $\sbis{\rm PFTr,\sqcup\sqcap}$ and $\sbis{\rm PF,\sqcup\sqcap}$ are also incomparable
with $\sbis{\rm PCTr}$ and $\sbis{\rm PTr}$. Finally, Figs.~\ref{fig:counterex_trace}(vi) and~(viii) show
that $\sbis{\rm PRTr}$ and $\sbis{\rm PRTr,\sqcup\sqcap}$ are incomparable with $\sbis{\rm PR}$ and
$\sbis{\rm PR,\sqcup\sqcap}$, Figs.~\ref{fig:counterex_trace}(ii) and~(vi) show that $\sbis{\rm
PFTr,\sqcup\sqcap}$ is incomparable with $\sbis{\rm PF}$, Figs.~\ref{fig:counterex_trace}(i) and~(v) show
that $\sbis{\rm PCTr}$ and $\sbis{\rm PCTr,\sqcup\sqcap}$ are incomparable with $\sbis{\rm PTr,dis}$, and
Figs.~\ref{fig:counterex_trace}(ii) and~(v) show that $\sbis{\rm PCTr,\sqcup\sqcap}$ is incomparable with
$\sbis{\rm PTr}$.

%
%
\section{Testing Equivalences for NPLTS Models}\label{sec:testing_equiv}
%
%

Testing equivalences consider the probability of two processes of performing computations along which the
same tests are passed. Tests specify which actions of a process are permitted at each step and, in this
setting, can be formalized as NPLTS models equipped with a success state. For the sake of simplicity, we
restrict ourselves to finite tests, each of which has finitely many states, finitely many outgoing
transitions from each state, an acyclic graph structure, and hence finitely many computations leading to
success.



	\begin{definition}

A nondeterministic and probabilistic test, NPT for short, is a finite NPLTS $\calt \! = \! (\! O, A, \!
\arrow{}{} \!\!)$ where $O$~contains a distinguished success state denoted by $\omega$ that has no outgoing
transitions. We say that a computation of~$\calt$ is successful iff its last state is $\omega$.
\fullbox

	\end{definition}

	\begin{definition}

Let $\call = (S, A, \! \arrow{}{} \!)$ be an NPLTS and $\calt = (O, A, \! \arrow{}{\calt} \!)$ be an NPT.
The interaction system of $\call$ and $\calt$ is the NPLTS $\cali(\call, \calt) = \call \pco{} \calt$ where:

		\begin{itemize}

\item Every element $(s, o) \in S \times O$ is called a configuration and is said to be successful iff $o =
\omega$.

\item A computation of $\cali(\call, \calt)$ is said to be successful iff its last configuration is
successful. Given $s \in S$, $o \in O$, and $\calz \in \ms{Res}(s, o)$, we denote by $\calsc(z_{s, o})$ the
set of successful computations from the state $z_{s, o}$ of $\calz$ corresponding to the configuration $(s,
o)$ of $\cali(\call, \calt)$.
\fullbox

		\end{itemize}

	\end{definition}

Due to the possible presence of equally labeled transitions departing from the same state, there is not
necessarily a single probability value with which an NPLTS passes a test. Thus, given two states $s_{1}$
and~$s_{2}$ of the NPLTS under test and the initial state $o$ of the test, we need to compute the
probability of performing a successful computation from the two configurations $(s_{1}, o)$ and $(s_{2}, o)$
in every maximal resolution of the interaction system. One option is comparing, for the two configurations,
\textit{only the extremal values of these success probabilities} over all maximal resolutions of the
interaction system. An alternative option is comparing \textit{all the success probabilities} and requiring
that for each maximal resolution of either configuration there is a matching maximal resolution of the other
configuration.

	\begin{definition}\label{def:ptesupinf}

(\emph{Probabilistic $\sqcup\sqcap$-testing equivalence} -- $\sbis{\textrm{PTe-}\sqcup\sqcap}$ --
\cite{YL92,JY95,Seg96,DGHM08}) \\
$s_{1} \sbis{\textrm{PTe-}\sqcup\sqcap} s_{2}$ iff for every NPT $\calt = (O, A, \! \arrow{}{\calt} \!)$
with initial state $o \in O$:
\cws{8}{\begin{array}{rcl}
\bigsqcup\limits_{\calz_{1} \in \ms{Res}_{\rm max}(s_{1}, o)} \ms{prob}(\calsc(z_{s_{1}, o})) & \!\!\! = 
\!\!\! & \bigsqcup\limits_{\calz_{2} \in \ms{Res}_{\rm max}(s_{2}, o)} \ms{prob}(\calsc(z_{s_{2}, o}))
\\[0.4cm]
\bigsqcap\limits_{\calz_{1} \in \ms{Res}_{\rm max}(s_{1}, o)} \ms{prob}(\calsc(z_{s_{1}, o})) & \!\!\! =
\!\!\! & \bigsqcap\limits_{\calz_{2} \in \ms{Res}_{\rm max}(s_{2}, o)} \ms{prob}(\calsc(z_{s_{2}, o})) \\
\end{array}}
\fullbox

	\end{definition}

	\begin{definition}\label{def:pteallexists}

(\emph{Probabilistic $\forall\exists$-testing equivalence} -- $\sbis{\textrm{PTe-}\forall\exists}$ --
\cite{BDL12}) \\
$s_{1} \sbis{\textrm{PTe-}\forall\exists} s_{2}$ iff for every NPT $\calt = (O, A, \! \arrow{}{\calt} \!)$
with initial state $o \in O$ it holds that for each \linebreak $\calz_{1} \in \ms{Res}_{\rm max}(s_{1}, o)$
there exists $\calz_{2} \in \ms{Res}_{\rm max}(s_{2}, o)$ such that:
\cws{0}{\ms{prob}(\calsc(z_{s_{1}, o})) \: = \: \ms{prob}(\calsc(z_{s_{2}, o}))}
and symmetrically for each $\calz_{2} \in \ms{Res}_{\rm max}(s_{2}, o)$.
\fullbox

	\end{definition}

\begin{figure}[thb]

\centerline{\includegraphics{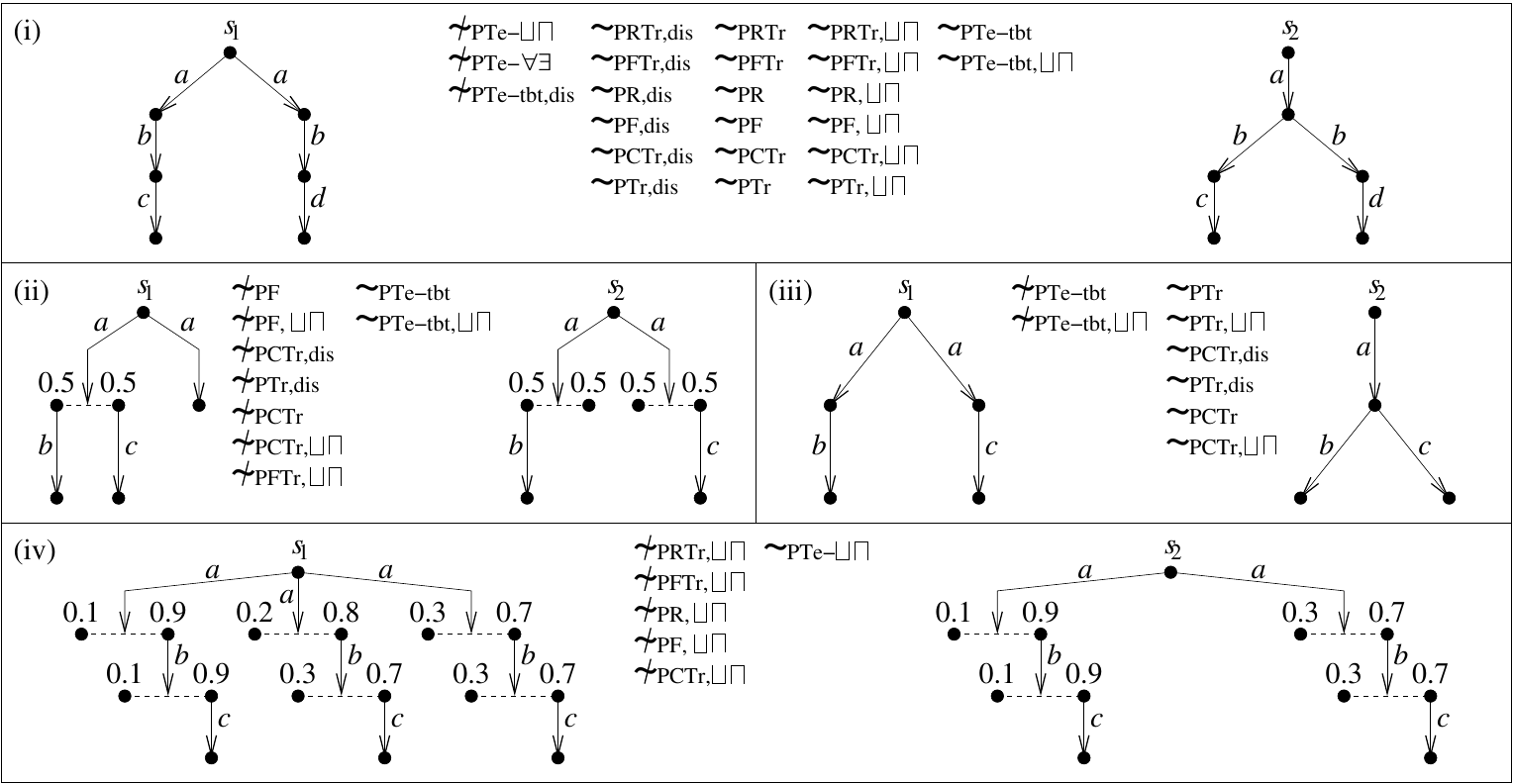}}
\caption{Counterexamples for strict inclusion and incomparability of the testing equivalences}
\label{fig:counterex_testing}

	\end{figure}

Neither $\sbis{\textrm{PTe-}\sqcup\sqcap}$ nor $\sbis{\textrm{PTe-}\forall\exists}$ is backward compatible
with the testing equivalence defined in~\cite{DH84} for fully nondeterministic processes. For instance,
Fig.~\ref{fig:counterex_testing}(i) shows two such processes related by classical testing equivalence that
are distinguished by $\sbis{\textrm{PTe-}\sqcup\sqcap}$ and $\sbis{\textrm{PTe-}\forall\exists}$. The reason
of the higher discriminating power of the latter two equivalences arises from the presence of probabilistic
choices within tests, which results in the capability of making copies of the process under
test~\cite{Abr87} and hence in an unrealistic estimation of success probabilities~\cite{GA10}. In order to
counterbalance this strong discriminating power, as illustrated in~\cite{BDL12} the idea is to consider
\emph{success probabilities in a trace-by-trace fashion} rather than on entire resolutions. Since traces
come again into play, the idea can be implemented in three different ways by following the three approaches
used in Sect.~\ref{sec:trace_equiv}.

In the following, given a state $s$ of an NPLTS, a state $o$ of an NPT, and a trace $\alpha \in A^{*}$, we
denote by $\ms{Res}_{{\rm max}, \calc, \alpha}(s, o)$ the set of resolutions $\calz \in \ms{Res}_{\rm
max}(s, o)$ such that $\calccc(z_{s, o}, \alpha) \neq \emptyset$, i.e., the maximal resolutions of $z_{s,
o}$ having at least one completed $\alpha$-compatible computation. Moreover, for each such resolution
$\calz$, we denote by $\calscc(z_{s, o}, \alpha)$ the set of successful $\alpha$-compatible computations
from $z_{s, o}$.

	\begin{definition}\label{def:ptetbtdis}

(\emph{Probabilistic trace-by-trace-distribution testing equivalence} -- $\sbis{\textrm{PTe-tbt,dis}}$) \\
$s_{1} \sbis{\textrm{PTe-tbt,dis}} s_{2}$ iff for every NPT $\calt = (O, A, \! \arrow{}{\calt} \!)$ with
initial state $o \in O$ it holds that for each \linebreak $\calz_{1} \in \ms{Res}_{\rm max}(s_{1}, o)$ there
exists $\calz_{2} \in \ms{Res}_{\rm max}(s_{2}, o)$ such that \underline{for all $\alpha \in A^{*}$} it
holds that $\calccc(z_{s_{1}, o}, \alpha) \neq \emptyset$ implies $\calccc(z_{s_{2}, o}, \alpha) \neq
\emptyset$ and:
\cws{0}{\ms{prob}(\calscc(z_{s_{1}, o}, \alpha)) \: = \: \ms{prob}(\calscc(z_{s_{2}, o}, \alpha))}
and symmetrically for each $\calz_{2} \in \ms{Res}_{\rm max}(s_{2}, o)$.
\fullbox

	\end{definition}

	\begin{definition}\label{def:ptetbt}

(\emph{Probabilistic trace-by-trace testing equivalence} -- $\sbis{\textrm{PTe-tbt}}$ -- \cite{BDL12}) \\
$s_{1} \sbis{\textrm{PTe-tbt}} s_{2}$ iff for every NPT $\calt = (O, A, \! \arrow{}{\calt} \!)$ with initial
state $o \in O$ and \underline{for all $\alpha \in A^{*}$} it holds that for each $\calz_{1} \in
\ms{Res}_{{\rm max}, \calc, \alpha}(s_{1}, o)$ there exists $\calz_{2} \in \ms{Res}_{{\rm max}, \calc,
\alpha}(s_{2}, o)$ such that:
\cws{0}{\ms{prob}(\calscc(z_{s_{1}, o}, \alpha)) \: = \: \ms{prob}(\calscc(z_{s_{2}, o}, \alpha))}
and symmetrically for each $\calz_{2} \in \ms{Res}_{{\rm max}, \calc, \alpha}(s_{2}, o)$.
\fullbox

	\end{definition}

	\begin{definition}\label{def:ptetbtsupinf}

(\emph{Probabilistic $\sqcup\sqcap$-trace-by-trace testing equivalence} --
$\sbis{\textrm{PTe-tbt},\sqcup\sqcap}$) \\
$s_{1} \sbis{\textrm{PTe-tbt},\sqcup\sqcap} s_{2}$ iff for every NPT $\calt = (O, A, \! \arrow{}{\calt} \!)$
with initial state $o \in O$ and for all $\alpha \in A^{*}$ it holds that $\ms{Res}_{{\rm max}, \calc,
\alpha}(s_{1}, o) \neq \emptyset$ iff $\ms{Res}_{{\rm max}, \calc, \alpha}(s_{2}, o) \neq \emptyset$ and:
\cws{8}{\begin{array}{rcl}
\bigsqcup\limits_{\calz_{1} \in \ms{Res}_{{\rm max}, \calc, \alpha}(s_{1}, o)} \ms{prob}(\calscc(z_{s_{1},
o}, \alpha)) & \!\!\! = \!\!\! & \bigsqcup\limits_{\calz_{2} \in \ms{Res}_{{\rm max}, \calc, \alpha}(s_{2},
o)} \ms{prob}(\calscc(z_{s_{2}, o}, \alpha)) \\[0.4cm]
\bigsqcap\limits_{\calz_{1} \in \ms{Res}_{{\rm max}, \calc, \alpha}(s_{1}, o)} \ms{prob}(\calscc(z_{s_{1},
o}, \alpha)) & \!\!\! = \!\!\! & \bigsqcap\limits_{\calz_{2} \in \ms{Res}_{{\rm max}, \calc, \alpha}(s_{2},
o)} \ms{prob}(\calscc(z_{s_{2}, o}, \alpha)) \\
\end{array}}
\fullbox

	\end{definition}

While only $\sbis{\textrm{PTe-tbt}}$ and $\sbis{\textrm{PTe-tbt},\sqcup\sqcap}$ are backward compatible with
the testing equivalence defined in~\cite{DH84} for fully nondeterministic processes -- which we denote by
$\sbis{\rm Te,fnd}$ (see Fig.~\ref{fig:counterex_testing}(i) for the counterexamples) -- all the five
testing equivalences defined above are backward compatible with the testing equivalence defined
in~\cite{CDSY99} for fully probabilistic processes -- which we denote by $\sbis{\rm Te,fpr}$.

	\begin{theorem}\label{thm:testing_compat}

It holds that:

		\begin{enumerate}

\item $\sbis{\textrm{PTe-tbt}} \: = \: \sbis{\textrm{PTe-tbt},\sqcup\sqcap} \: = \: \sbis{\rm Te,fnd}$ over
fully nondeterministic NPLTS models.

\item $\sbis{\textrm{PTe-}\sqcup\sqcap} \: = \: \sbis{\textrm{PTe-}\forall\exists} \: = \:
\sbis{\textrm{PTe-tbt,dis}} \: = \: \sbis{\textrm{PTe-tbt}} \: = \: \sbis{\textrm{PTe-tbt},\sqcup\sqcap} \:
= \: \sbis{\rm Te,fpr}$ over fully probabilistic NPLTS \linebreak models.
\fullbox

		\end{enumerate}

	\end{theorem}

We now investigate the relationships of the five testing equivalences among themselves (first two properties
below) and with the eighteen trace-based equivalences (last three properties below). It turns out that
$\sbis{\textrm{PTe-}\forall\exists}$ and $\sbis{\textrm{PTe-tbt,dis}}$ perform exactly the same
identifications.  Unlike the fully nondeterministic spectrum -- where the testing semantics coincides with
the failure semantics when all actions are observable~\cite{DeN87} -- here $\sbis{\textrm{PTe-tbt,dis}}$ is
finer than $\sbis{\rm PFTr,dis}$ while $\sbis{\textrm{PTe-tbt}}$ and $\sbis{\textrm{PTe-tbt},\sqcup\sqcap}$
are coarser than $\sbis{\rm PF}$ and $\sbis{\rm PF,\sqcup\sqcap}$, respectively. In contrast,
$\sbis{\textrm{PTe-}\sqcup\sqcap}$ has no inclusion relationship with the failure semantics.

	\begin{theorem}\label{thm:testing_results}

It holds that:

		\begin{enumerate}

\item $\sbis{\textrm{PTe-}\forall\exists} \: \subseteq \: \sbis{\textrm{PTe-}\sqcup\sqcap} \: \subseteq \:
\sbis{\textrm{PTe-tbt},\sqcup\sqcap}$.

\item $\sbis{\textrm{PTe-}\forall\exists} \: = \: \sbis{\textrm{PTe-tbt,dis}} \: \subseteq \:
\sbis{\textrm{PTe-tbt}} \: \subseteq \: \sbis{\textrm{PTe-tbt},\sqcup\sqcap}$.

\item $\sbis{\textrm{PTe-tbt,dis}} \: \subseteq \: \sbis{\rm PRTr,dis}$.

\item $\sbis{\rm PF} \: \subseteq \: \sbis{\textrm{PTe-tbt}} \: \subseteq \: \sbis{\rm PTr}$.

\item $\sbis{\rm PF,\sqcup\sqcap} \: \subseteq \: \sbis{\textrm{PTe-tbt},\sqcup\sqcap} \: \subseteq \:
\sbis{\rm PTr,\sqcup\sqcap}$.
\fullbox

		\end{enumerate}

	\end{theorem}

All the inclusions above are strict, as shown in Figs.~\ref{fig:counterex_trace}(i) and~(ii) and
Figs.~\ref{fig:counterex_testing}(i) to~(iii). It is worth noting the isolation of
$\sbis{\textrm{PTe-}\sqcup\sqcap}$, which is incomparable with $\sbis{\rm PRTr,dis}$, $\sbis{\rm PFTr,dis}$,
$\sbis{\rm PR,dis}$, $\sbis{\rm PF,dis}$, $\sbis{\rm PCTr,dis}$, $\sbis{\rm PTr,dis}$, $\sbis{\rm PRTr}$,
$\sbis{\rm PFTr}$, $\sbis{\rm PR}$, $\sbis{\rm PF}$, $\sbis{\rm PCTr}$, $\sbis{\rm PTr}$, and
$\sbis{\textrm{PTe-tbt}}$, as shown in Fig.~\ref{fig:counterex_trace}(ii) and
Fig.~\ref{fig:counterex_testing}(i), and with $\sbis{\rm PRTr,\sqcup\sqcap}$, $\sbis{\rm
PFTr,\sqcup\sqcap}$, $\sbis{\rm PR,\sqcup\sqcap}$, $\sbis{\rm PF,\sqcup\sqcap}$, and $\sbis{\rm
PCTr,\sqcup\sqcap}$, as shown in Fig.~\ref{fig:counterex_trace}(i) and Fig.~\ref{fig:counterex_testing}(iv).
Furthermore, $\sbis{\textrm{PTe-tbt}}$ and $\sbis{\textrm{PTe-tbt},\sqcup\sqcap}$ are incomparable with
$\sbis{\rm PRTr}$, $\sbis{\rm PR}$, $\sbis{\rm PRTr,\sqcup\sqcap}$, and $\sbis{\rm PR,\sqcup\sqcap}$, as
shown in Fig.~\ref{fig:counterex_trace}(vii), and with $\sbis{\rm PCTr,dis}$, $\sbis{\rm PTr,dis}$,
$\sbis{\rm PCTr}$, and $\sbis{\rm PCTr,\sqcup\sqcap}$, as shown in Figs.~\ref{fig:counterex_testing}(ii)
and~(iii). Finally, Figs.~\ref{fig:counterex_trace}(ii) and~\ref{fig:counterex_testing}(ii) show that
$\sbis{\textrm{PTe-tbt}}$ is also incomparable with $\sbis{\rm PFTr,\sqcup\sqcap}$ and $\sbis{\rm
PF,\sqcup\sqcap}$, while Figs.~\ref{fig:counterex_trace}(ii) and~\ref{fig:counterex_testing}(iii) show that
$\sbis{\textrm{PTe-tbt},\sqcup\sqcap}$ is also incomparable with $\sbis{\rm PTr}$.

%
%
\section{Bisimulation Equivalences for NPLTS Models}\label{sec:bisim_equiv}
%
%

Bisimulation equivalences capture the ability of two processes of mimicking each other's behavior stepwise.
Similar to the trace-based case, given two states there are three different approaches to the definition of
these bisimilarities, each following the style of~\cite{LS91} based on equivalence relations. The first
approach is to match transitions on the basis of \emph{class distributions}, which means that for each
transition of one of the two states there must exist an equally labeled transition of the other state such
that, \emph{for every equivalence class}, the two transitions have the same probability of reaching a state
in that class. In other words, matching transitions of the two states are related by the fully probabilistic
version of bisimilarity (fully matching transitions). The second approach is to consider \emph{a single
equivalence class at a time}, i.e., to anticipate the quantification over classes. In this way, a transition
departing from one of the two states is allowed to be matched, with respect to the probabilities of reaching
different classes, by several different transitions departing from the other state (partially matching
transitions). The third approach is to compare only the \emph{extremal probabilities} of reaching each class
over all possible transitions labeled with a certain action (max-min-matching transition sets).

Unlike~\cite{LS91}, we will consider \textit{groups of equivalence classes} rather than individual
equivalence classes. This does not change the discriminating power in the case of the first approach, while
it increases the discriminating power thereby resulting in desirable logical characterizations in the case
of the other two approaches~\cite{DMRS08,TDZ11,SZG11,BDL13c}. Given an NPLTS $(S, A, \! \arrow{}{} \!)$ and
a distribution $\cald \in \ms{Distr}(S)$, in the following we let $\cald(S') = \sum_{s \in S'} \cald(s)$ for
$S' \subseteq S$. Moreover, given an equivalence relation $\calb$ over~$S$ and a group of equivalence
classes $\calg \in 2^{S / \calb}$, we also let $\bigcup \calg = \bigcup_{C \in \calg} C$.

	\begin{definition}\label{def:pbdis}

(\emph{Probabilistic group-distribution bisimilarity} -- $\sbis{\rm PB,dis}$ -- \cite{SL94}) \\
$s_{1} \sbis{\rm PB,dis} s_{2}$ iff $(s_{1}, s_{2})$ belongs to the largest probabilistic group-distribution
bisimulation. An equivalence relation $\calb$ over $S$ is a \emph{probabilistic group-distribution
bisimulation} iff, whenever $(s_{1}, s_{2}) \in \calb$, then for each $s_{1} \arrow{a}{} \cald_{1}$ there
exists $s_{2} \arrow{a}{} \cald_{2}$ such that \underline{for all $\calg \in 2^{S / \calb}$} it holds that
$\cald_{1}(\bigcup \calg) = \cald_{2}(\bigcup \calg)$.
\fullbox

	\end{definition}

	\begin{definition}\label{def:pb}

(\emph{Probabilistic bisimilarity} -- $\sbis{\rm PB}$ -- \cite{BDL13c}) \\
$s_{1} \sbis{\rm PB} s_{2}$ iff $(s_{1}, s_{2})$ belongs to the largest probabilistic bisimulation. An
equivalence relation $\calb$ over~$S$ is a \emph{probabilistic bisimulation} iff, whenever $(s_{1}, s_{2})
\in \calb$, then \underline{for all $\calg \in 2^{S / \calb}$} it holds that for each $s_{1} \arrow{a}{}
\cald_{1}$ there exists $s_{2} \arrow{a}{} \cald_{2}$ such that $\cald_{1}(\bigcup \calg) =
\cald_{2}(\bigcup \calg)$.
\fullbox

	\end{definition}

	\begin{figure}[thb]

\centerline{\includegraphics{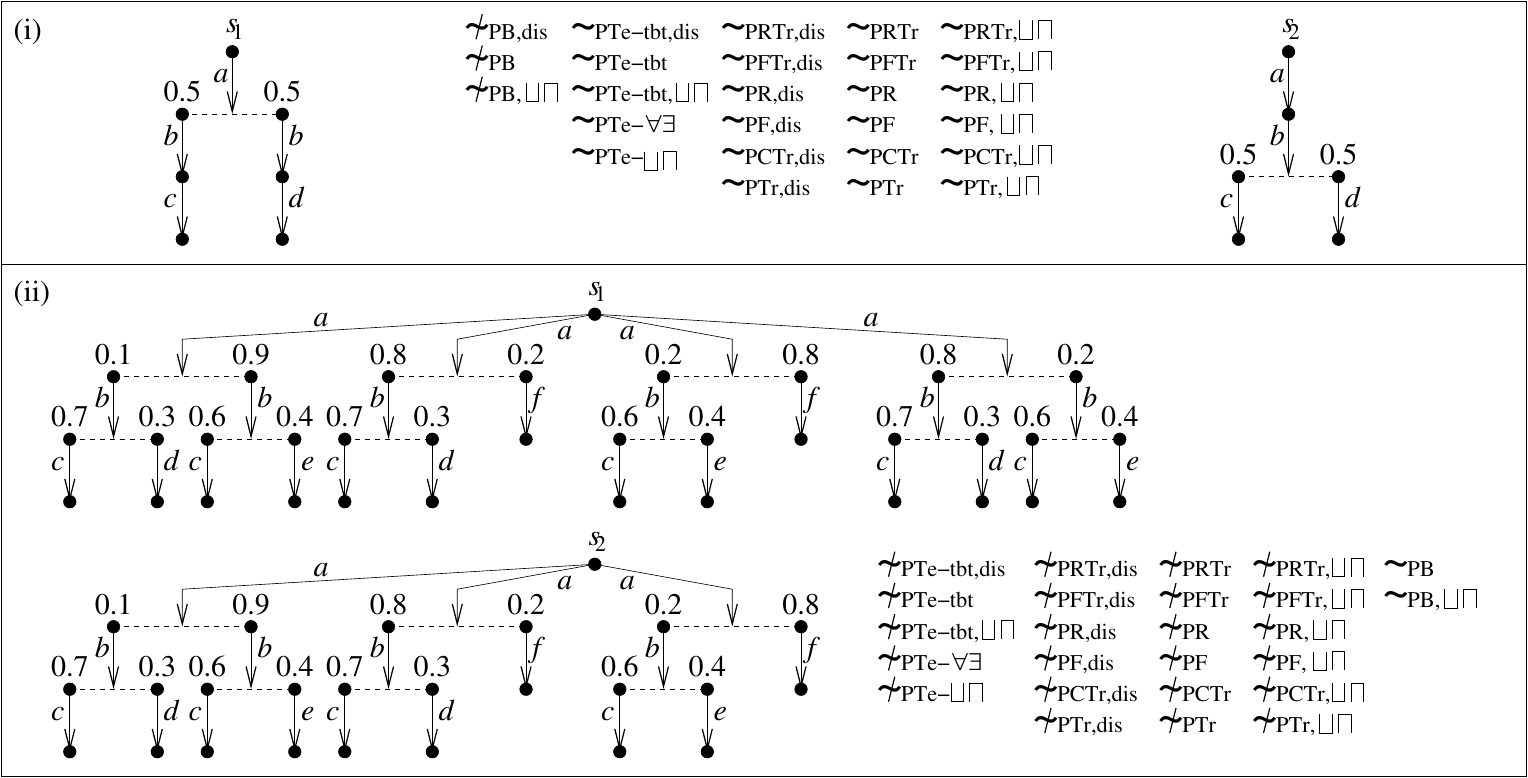}}
\caption{Counterexamples for strict inclusion and incomparability of the bisimulation equivalences}
\label{fig:counterex_bisim}

	\end{figure}

	\begin{definition}\label{def:pbsupinf}

(\emph{Probabilistic $\sqcup\sqcap$-bisimilarity} -- $\sbis{\rm PB,\sqcup\sqcap}$ -- \cite{BDL13c}) \\
$s_{1} \sbis{\rm PB,\sqcup\sqcap} s_{2}$ iff $(s_{1}, s_{2})$ belongs to the largest probabilistic
$\sqcup\sqcap$-bisimulation. An equivalence relation $\calb$ over $S$ is a \emph{probabilistic
$\sqcup\sqcap$-bisimulation} iff, whenever $(s_{1}, s_{2}) \in \calb$, then for all $\calg \in 2^{S /
\calb}$ and $a \in A$ \linebreak it holds that $s_{1} \arrow{a}{} \!$ iff $s_{2} \arrow{a}{} \!$ and:
\cws{10}{\begin{array}{rcl}
\bigsqcup\limits_{s_{1} \arrow{a}{} \cald_{1}} \cald_{1}(\bigcup \calg) & \!\!\! = \!\!\! &
\bigsqcup\limits_{s_{2} \arrow{a}{} \cald_{2}} \cald_{2}(\bigcup \calg) \\[0.4cm]
\bigsqcap\limits_{s_{1} \arrow{a}{} \cald_{1}} \cald_{1}(\bigcup \calg) & \!\!\! = \!\!\! &
\bigsqcap\limits_{s_{2} \arrow{a}{} \cald_{2}} \cald_{2}(\bigcup \calg) \\
\end{array}}
\fullbox

	\end{definition}

The three bisimulation equivalences defined above are all backward compatible with the bisimulation
equivalences respectively defined in~\cite{HM85} for fully nondeterministic processes -- which we denote by
$\sbis{\rm B,fnd}$ -- and in~\cite{GJS90} for fully probabilistic processes -- which we denote by $\sbis{\rm
B,fpr}$.

	\begin{theorem}\label{thm:bisim_compat}

It holds that:

		\begin{enumerate}

\item $\sbis{\rm PB,dis} \: = \: \sbis{\rm PB} \: = \: \sbis{\rm PB,\sqcup\sqcap} \: = \: \sbis{\rm B,fnd}$
over fully nondeterministic NPLTS models.

\item $\sbis{\rm PB,dis} \: = \: \sbis{\rm PB} \: = \: \sbis{\rm PB,\sqcup\sqcap} \: = \: \sbis{\rm B,fpr}$
over fully probabilistic NPLTS models.
\fullbox

		\end{enumerate}

	\end{theorem}

We now investigate the relationships of the three bisimulation equivalences among themselves (first property
below) and with the five testing equivalences and the eighteen trace-based equivalences (second property
below).

	\begin{theorem}\label{thm:bisim_results}

It holds that:

		\begin{enumerate}

\item $\sbis{\rm PB,dis} \: \subseteq \: \sbis{\rm PB} \: \subseteq \: \sbis{\rm PB,\sqcup\sqcap}$.

\item $\sbis{\rm PB,dis} \: \subseteq \: \sbis{\textrm{PTe-tbt,dis}}$.
\fullbox

		\end{enumerate}

	\end{theorem}

All the inclusions above are strict, as shown in Figs.~\ref{fig:counterex_trace}(i) and~(ii) and
Fig.~\ref{fig:counterex_bisim}(i). It is worth noting the isolation of $\sbis{\rm PB}$ and $\sbis{\rm
PB,\sqcup\sqcap}$, which are incomparable with all the five testing equivalences and all the eighteen
trace-based equivalences, as shown in Figs.~\ref{fig:counterex_bisim}(i) and~(ii).

%
%
\section{Conclusion}\label{sec:concl}
%
%

\begin{figure}[thb]

\centerline{\includegraphics{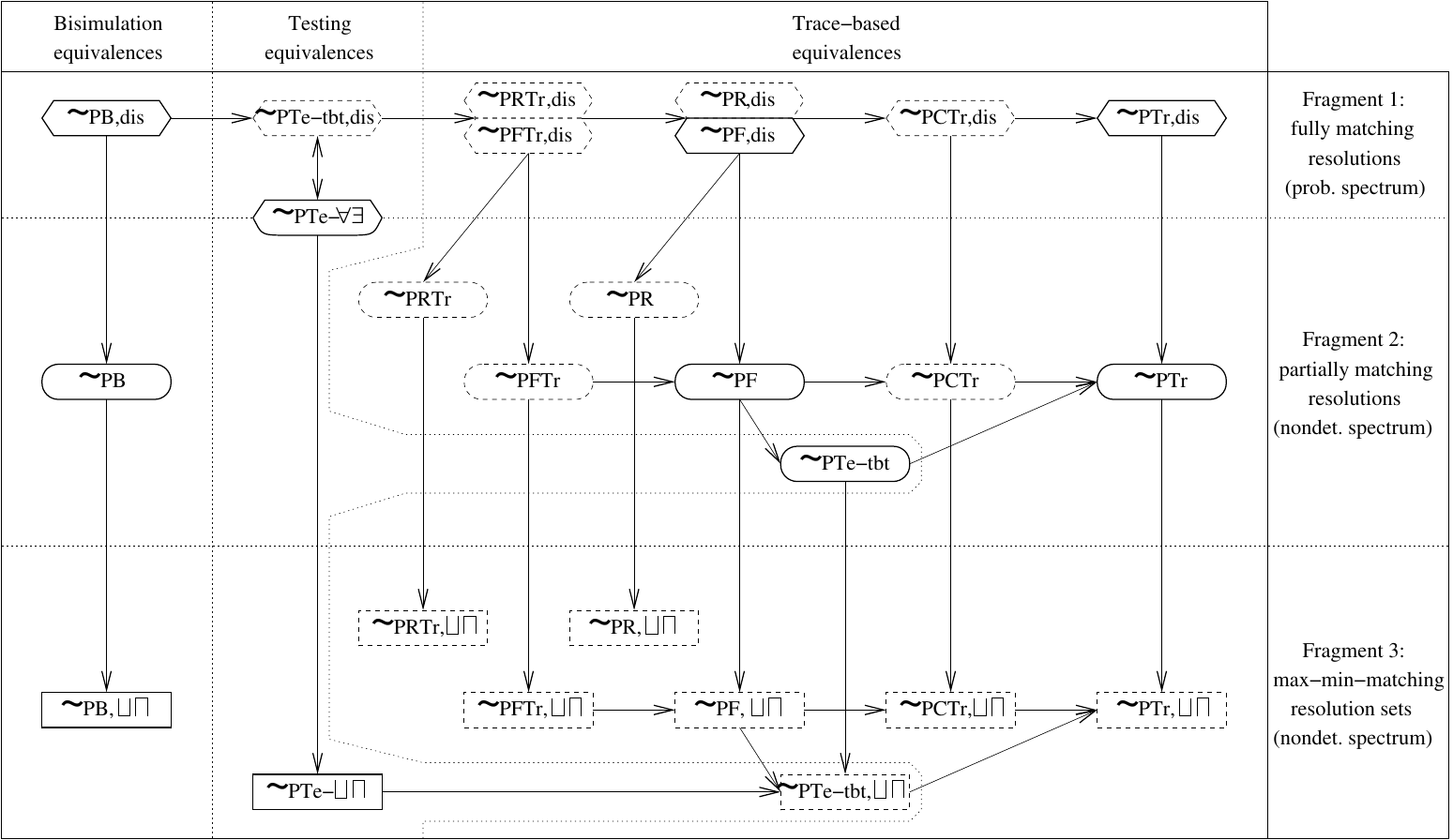}}
\caption{Spectrum of strong behavioral equivalences for NPLTS models (deterministic schedulers)}
\label{fig:spectrum}

	\end{figure}

We have studied the relationships among the equivalences that stem from three significantly different
approaches to the definition of behavioral relations for NPLTS models. The specificity of the three
approaches is determined by the way they deal with the probabilities associated with the resolutions of
nondeterminism. For each approach, we have considered the families of strong trace-based, testing, and
bisimulation equivalences under deterministic schedulers. The relationships among the equivalences for
finitely-branching NPLTS models are summarized in Fig.~\ref{fig:spectrum}. In the spectrum, the absence of
(chains of) arrows represents incomparability, adjacency of boxes within the same fragment and double arrows
connecting boxes of different fragments indicate coincidence, and single arrows stand for the
strictly-more-discriminating-than relation.

Continuous hexagonal boxes contain equivalences studied in the last twenty years~\cite{SL94,Seg95b,Seg96},
which compare probability distributions of all equivalence-specific events. In contrast, continuous rounded
boxes contain equivalences assigning a weaker role to schedulers that have been recently introduced
in~\cite{DMRS08,TDZ11,SZG11,BDL12,BDL13c}, which compare separately the probabilities of individual
equivalence-specific events. Continuous rectangular boxes instead contain old
equivalences~\cite{YL92,JY95,Seg96,DGHM08} and new equivalences~\cite{BDL13c} based on extremal
probabilities. The only hybrid box is the one containing $\sbis{\textrm{PTe-}\forall\exists}$, as this
equivalence does not follow any of the three definitional approaches. Finally, dashed boxes contain
equivalences defined for the first time in this paper to better assess the different impact of the
approaches themselves.

Figure~\ref{fig:spectrum} evidences that the top fragment of the spectrum collapses several equivalences,
whilst the middle fragment and the bottom fragment do not. Indeed, like in the spectrum for fully
probabilistic processes~\cite{JS90,HT92}, we have that the top variants of ready-trace and failure-trace
equivalences and of readiness and failure equivalences respectively induce the same identifications. In
contrast, the more liberal variants in the middle fragment and the bottom fragment, which guarantee a higher
degree of flexibility in determining the matching resolutions and are in general coarser, do not flatten the
specificity of the intuition behind the original definition of the behavioral equivalences for LTS models.
Therefore, those two fragments preserve much of the original spectrum of~\cite{Gla01} for fully
nondeterministic processes. We finally stress again the isolation of $\sbis{\rm PB}$, $\sbis{\rm
PB,\sqcup\sqcap}$, $\sbis{\textrm{PTe-}\sqcup\sqcap}$, $\sbis{\rm PRTr}$, $\sbis{\rm PR}$, $\sbis{\rm
PRTr,\sqcup\sqcap}$, and $\sbis{\rm PR,\sqcup\sqcap}$.

As future work, we intend first of all to enrich the spectrum with simulation equivalences. Secondly, we
plan to address how the spectrum changes if randomized schedulers are used. Thirdly, we would like to
investigate the spectrum of weak behavioral equivalences, for which the choice of randomized schedulers is
more appropriate. Finally, it would be interesting to compare the discriminating power of the various
equivalences after defining them more abstractly on a parametric model. A suitable framework might be that
of \ultras~\cite{BDL13a}, as it has been shown to encompass trace, testing, and bisimulation equivalences
for models such as labeled transition systems, discrete-/continuous-time Markov chains, and
discrete-/continuous-time Markov decision processes without/with internal nondeterminism.

\bibliographystyle{eptcs}
\bibliography{qapl2013}

\end{document}

%% file: Commands.tex
\def\ms#1{\null\ifmmode\mathord{\mathcode`-="702D\it #1\mathcode`\-="2200}%
	\else$\mathord{\mathcode`-="702D\it #1\mathcode`\-="2200}$\fi}

\newcommand{\cws}[2]
	{\\ \centerline{$#2$} \\[-#1pt]}

\newlength{\spacelen}

\newcommand{\bibtrick}[1]
	{}

\newcommand{\calb}
        {\mathcal{B}}

\newcommand{\calc}
        {\mathcal{C}}

\newcommand{\calcc}
        {\mathcal{CC}}

\newcommand{\calccc}
        {\mathcal{CCC}}

\newcommand{\cald}
        {\mathcal{D}}

\newcommand{\calfcc}
        {\mathcal{FCC}}

\newcommand{\calftcc}
        {\mathcal{FTCC}}

\newcommand{\calg}
        {\mathcal{G}}

\newcommand{\cali}
        {\mathcal{I}}

\newcommand{\call}
        {\mathcal{L}}

\newcommand{\calrcc}
        {\mathcal{RCC}}

\newcommand{\calrtcc}
        {\mathcal{RTCC}}

\newcommand{\calsc}
        {\mathcal{SC}}

\newcommand{\calscc}
        {\mathcal{SCC}}

\newcommand{\calt}
        {\mathcal{T}}

\newcommand{\calz}
        {\mathcal{Z}}

\newcommand{\natns}
	{\mathbb{N}}

\newcommand{\realns}
	{\mathbb{R}}

\newcommand{\step}[2]
        {\, {\auxstep\limits^{#1}}_{#2} \,}
\newcommand{\auxstep}
	{\mathop{-\hspace{-0.15cm}\mapsto}}

\newcommand{\arrow}[2]
        {\, {\auxarrow\limits^{#1}}_{#2} \,}
\newcommand{\auxarrow}
	{\mathop{\longrightarrow}}

\newcommand{\sbis}[1]
	{\sim_{#1}}

\newcommand{\pco}[1]
	{\mathop{\Vert_{#1}}}

\newcommand{\fullbox}
	{{\mbox{}\nolinebreak\hfill{$\rule{2mm}{2mm}$}}}

\newcommand{\ultras}
	{{\sc ULTraS}}


%% file: Environments.tex
\newtheorem{new_theorem}
	{Theorem}[section]

\newtheorem{new_definition}
	[new_theorem]{Definition}

\newtheorem{new_remark}
	[new_theorem]{Remark}

\newtheorem{new_example}
	[new_theorem]{Example}

\newtheorem{new_lemma}
	[new_theorem]{Lemma}

\newtheorem{new_proposition}
	[new_theorem]{Proposition}

\newtheorem{new_corollary}
	[new_theorem]{Corollary}

\newenvironment{definition}
	{\begin{new_definition}\rm}
	{\end{new_definition}}

\newenvironment{theorem}
	{\begin{new_theorem}\rm}
	{\end{new_theorem}}

